# The H$^+$ region Contribution to [C II] 158 Micron Emission

N. P. Abel
University of Kentucky,
Department of Physics and Astronomy,
Lexington, KY 40506
npabel2@uky.edu

# Abstract

The [C II] 158 μm line is an important emission line diagnostic in Photodissociation Regions (PDRs), but this emission line can also emerge from ionized gas. This work calculates the contribution of [C II] emission from ionized gas over a wide range of parameter space by considering the simplified case of an $H^+$ region and PDR in pressure equilibrium. Additionally, these calculations also predict the strong correlation observed between [N II] 205μm emission and [C II] discussed by previous authors. Overall, the results of these calculations have wide-ranging applications to the interpretation of [C II] emission in astrophysical environments.

**Keywords:** line-formation, H II regions, PDRs, equation of state

# 1 Introduction

The [C II] $^2P_{3/2}$ → $^2P_{1/2}$ emission line at 158 μm (henceforth referred to as [C II]) is an important diagnostic in photodissociation regions (PDRs). The [C II] line is one of the brightest lines in the infrared (IR). The ISO mission observed the [C II] line in many star-forming regions (see, for instance, Malhotra et al. 2001; Liseau, Justtanont, & Tielens 2006). Both SOFIA and Herschel will also observe the [C II] line in a wide range of astrophysical environments. A correct interpretation of [C II] emission, therefore, is vital to enhance the scientific return of these missions.

This paper focuses on the contribution of [C II] emission from $H^+$ regions over a wide range of observed $H^+$ region conditions. Section 2 describes the current state of this problem, Section 3 gives the details of our calculations, with results shown in Section 4. A series of conclusions is given in Section 5.

# 2 Physics and Diagnostic Capabilities of [C II] Emission

The [C II] emission line plays an important role in probing the physical conditions in PDRs. Physically, the [C II] line is an important coolant in PDRs (Tielens & Hollenbach 1985). Additionally, there are strong correlations between [C II] emission and emission lines, such as [O I] 63, 146 μm, [C I] 370, 610 μm, CO (*J*=1-0), and the far infrared flux (FIR) (Malhotra et al. 2001).

## 2.1 [C II] Formation and Diagnostic Properties

Carbon's abundance and first ionization potential makes the [C II] line readily observable. Carbon is usually the fourth most abundant element, with only H, He, and O being more abundant. Of these four, only carbon remains ionized in a PDR, since only carbon has a first ionization potential less than 13.6 eV (a PDR is defined as the region where 6 – 13.6 eV photons dominate the physical processes – see Tielens & Hollenbach 1985). The large abundance of $C^+$ means the electron density ($n_e$) approximately equals the $C^+$ density in a PDR, with a small contribution from $H^+$, $S^+$, $Si^+$, $Mg^+$, and $Fe^+$. Carbon remains in the form of $C^+$ until after $H_2$ fully forms, when most photons capable of ionizing $C^0$ have been absorbed and carbon makes the transition from $C^+ \rightarrow C^0 \rightarrow CO$.

[C II] emission is produced by collisional excitation followed by radiative decay. Collisions of $C^+$ with $H^0$, $H_2$, and $e^-$ populate the $^2P_{3/2}$ level, and then decays into the $^2P_{1/2}$ state emit radiation at 157.6 μm. Radiative decay is efficient as long as the $n(H)$, $n(H_2)$, and $n_e$ densities are below their critical densities ($n_c$) of $3\times10^3$, $5\times10^3$, and 50 cm$^{-3}$ respectively (Genzel 1991), and as long as the energy of the colliding partner exceeds the $^2P_{3/2} \rightarrow {}^2P_{1/2}$ energy difference of 92 K. When the density is above $n_c$, the $^2P_{3/2}$ level is preferentially de-excited through collisions, and [C II] emission is suppressed.

Correlations between [C II] emission and other emission lines are invaluable in determining the physical conditions in PDRs. The general problem is to determine what hydrogen density, $n_H$, and 6 and 13.6 eV radiation flux (normally referred to as $G_0$, where a $1G_0 = 1.6\times10^{-3}$ erg cm$^{-2}$ s$^{-1}$; Habing 1968) reproduces the observed PDR emission line spectrum (see, for instance, Kaufman et al. 1999). The quantities $n_H$ and $G_0$ determine the chemical/thermal structure of the PDR (assuming a set of elemental abundances and grain model). These quantities can also be used to determine the mass, atomic and molecular gas temperature, filling factor, and number/radii of clouds along the line of sight (see Wolfire, Tielens, & Hollenbach 1990). Since [C II] emission is almost always observed in a PDR, observations of the intensity of [C II] are almost always used to constrain PDR models and determine physical conditions.

## 2.2 [C II] Emission in H$^+$ Regions

Although typically carbon is in the form of $C^{2+}$ in an H$^+$ region (see, for example, Figure 5 of Abel et al. (2005)), $C^+$ will also exist in the H$^+$ region. It is clear, for PDRs adjacent to H$^+$ regions, that the H$^+$ region must contribute something to the total [C II] emission. Theoretical calculations that predict the emission line spectrum of a PDR start at the ionization front and ignore the H$^+$ region. Since [C II] emission can come from both regions, predictions of [C II] emission must consider both regions. In order to use a standard PDR calculation to derive $n_H$ and $G_0$, the H$^+$ region contribution to the [C II] emission-line must be removed.

There are many approaches used to determine the H$^+$ region contribution to the [C II] emission. One approach is to do a single, self-consistent calculation of the H$^+$ region and PDR (Abel et al. 2005). Such a calculation determines the physical conditions of the H$^+$ region and PDR by assuming an equation of state (such as constant pressure). Another approach is to perform two separate calculations, one for the $H^+$ region and one for the PDR (Carral et al. 1994; van van den Ancker, Tielens, & Wesselius 2000). In this approach, the equation of state is included by increasing $n_H$ in the PDR, $n_{PDR}$, by a factor of $10^2$ relative to $n_H$ in the $H^+$ region, $n_{H^+}$. This is done to simulate constant pressure, where the temperature in the $H^+$ region and PDR is $10^4$ and $10^2$ K. Recent calculations by Kaufman, Hollenbach, & Wolfire (2006) have a similar approach, except instead of assuming a relative change in $n$ and $T$ of 2 dex, they take into account the differences in $T$ that arise at the PDR surface. The third approach is to use the observed correlation between [N II] 205μm emission (henceforth referred to as [N II] emission) and [C II] emission in the $H^+$ region. Petuchowski & Bennett (1993) and Heiles (1994), through observational data obtained with COBE and analytical considerations, have derived convenient formulas relating the [N II] emission to [C II]. However, these formulas were derived for the Warm Ionized Medium (WIM), and not the case of an $H^+$ region adjacent to a PDR.

A single calculation has many advantages over doing a separate calculation for the $H^+$ region and PDR. Theoretical calculations specialized for the $H^+$ region typically calculate the photoionization rate ($\Gamma$) by integrating the cross section ($\sigma$) for photoionization over the local continuum, $\Gamma = \int_{\nu_0}^{\infty} \frac{4\pi J_\nu}{h\nu} \sigma_\nu d\nu$ (s$^{-1}$). PDR calculations, for those species with ionization potentials less than 13.6 eV, will often use functions of the form $\Gamma = \alpha \exp(-bA_V)$, where α is the unshielded photoionization rate, $A_V$ is the visual extinction, and $b$ is a parameter used to take into account the increased extinction of dust in the UV (Le Teuff, Millar, & Markwick 2000). Separate calculation must assure $\Gamma$ is continuous, which requires that the treatment of grain physics and radiative transfer (ALI vs. escape probability) be identical. A separate calculation must take great care in assuring that the transmitted continuum emerging from the $H^+$ region is consistent with the initial conditions for the PDR. In a dynamical environment, the only way to treat the dynamics is to integrate the dynamical equations self-consistently across the $H^+$/PDR boundary. Treating constant pressure by assuming $n_{PDR} = 100 \times n_{H^+}$ is also not always accurate. Kaufman et al. (1999) show that the PDR surface temperature varies from 10 – 2000 K for $n_{PDR}$ between $10^2$ – $10^6$ cm$^{-3}$ and $G_0$ from 1 to $10^6$, meaning $n_{H^+}$ can only be a factor of five less than $n_{PDR}$. Abel et al. (2005) showed that, in the case of constant pressure, the relationship between $n_{H^+}$ and $n_{PDR}$ is approximately $n_{PDR} \approx (5\text{-}30) \times n_{H^+}$ for $n_{H^+} = 10 - 10^4$ cm$^{-3}$ and ionization parameter $U = 10^{-4} - 10^{-1}$ (where $U = \frac{\phi_H}{n_{H^+} c}$ is the dimensionless ratio of hydrogen ionizing flux to density). The PDR temperature does vary with depth, which under the constant pressure assumption means the density will also vary in the PDR. An isothermal or constant density PDR will obviously produce a different emission-line spectrum than a calculation that determines the temperature (and density) self-consistently (Roellig et al. in preparation; website http://hera.ph1.uni-koeln.de/~roellig). None of these problems appears in a self-consistent calculation.

## 3 Computational Details

The developmental version of the spectral synthesis code Cloudy, last described by Ferland et al. (1998), is used in the calculations. The details of the calculations are nearly identical to Abel et al. (2005). Therefore, the computational details are only reviewed here.

The atomic data used to calculate [C II] and [N II] 205 µm emission comes from several sources. For [C II], we use transition probabilities given in Froese Fischer (1983), while for [N II] we use Nussbaumer & Rusca (1979). We consider excitation of [C II] due to collisions with $e^-$, H, and $H_2$ and use collision strengths given in Wilson & Bell (2002) and Barinovs et al. (2005). For [N II] excitation due to electron collisions, we use collision strengths taken from Lennon & Burke (1994).

The molecular reaction network in Cloudy is discussed in Abel et al. (2005) and Shaw et al. (2005). The calculations include all stages of ionization for the lightest 30 elements. A complete model of the hydrogen atom is used in calculating the H I emission-line spectrum and ionization rates (Ferguson & Ferland 1997), as well as a complete model of the helium iso-electronic sequence (Porter et al. 2005).

The treatment of grain physics is described in van Hoof et al. (2004). A galactic ratio of visual extinction to hydrogen column density, $A_V/N(H)$, of $5\times10^{-22}$ mag cm$^2$, is used in all calculations. Grain size distributions for gas adjacent to $H^+$ regions, such as Orion (Cardelli et al. 1989) and starburst galaxies (Calzetti et al. 2000) tend to have a larger ratio of total to selective extinction (R) than observed in the ISM. A truncated MRN grain size distribution (Mathis et al. 1977) with $R = 5.5$, which reproduces the Orion extinction curve (Baldwin et al. 1991) is therefore used. The effects of PAHs are also included, with the same size distribution that is used in Bakes & Tielens (1994). PAHs are thought to exist mainly in $H^0$ regions (Giard et al. 1994), therefore the PAH abundance is scaled by the ratio of $H^0/H_{tot}$ ($n_C(PAH)/n_H = 3\times10^{-6} \times [n(H^0)/n(H_{tot})]$).

The assumed elemental abundances are an average set of abundances derived for the Orion Nebula. For the most important species, the abundances by number are He/H = 0.095, C/H= $3\times10^{-4}$; O/H= $4\times10^{-4}$, N/H= $7\times10^{-5}$, Ne/H= $6\times10^{-5}$, and Ar/H= $3\times10^{-6}$. The S/H ratio is taken to be $2\times10^{-6}$ and is based on observations of starburst galaxies by Verma et al. (2003). This is a factor of 5 lower than in Orion. Tests show that either assumed abundance does not change our results in any way. The complete set of assumed abundances can be found in Ferland (2002).

A wide range of H$^+$ regions are considered in the calculations. The $n_{H^+}$ density in the calculations vary from $10^1$ and $10^4$ cm$^{-3}$. The stellar continuum is defined by the ionization parameter ($U$) and stellar temperature ($T^*$). $U$ is the dimensionless ratio of hydrogen ionizing flux to density, and is a convenient way to characterize H$^+$ regions (Morisset 2004). The calculations presented here vary $U$ from $10^{-4}$ and $10^{-1}$. Both $n_{H^+}$ and $U$ were incremented by 1 dex. Three different stellar continuum sources are considered, those being Kurucz (1979) stellar atmosphere, the WMBasic supergiant atmosphere (Pauldrach, Hoffmann, & Lennon, M. 2001), and a Blackbody continuum. $T_*$ is allowed to vary from 30,000 to 50,000 K in increments of 5,000 K. All calculations stop at an $A_V$ of 100 magnitudes, ensuring that they extend deep into the molecular cloud. Overall, the combination of $n_{H^+}$, $U$, $T_*$, and stellar atmospheres constitute 240 separate calculations.

The H$^+$ region and PDR are dynamically linked by assuming constant pressure. Constant pressure is a first approximation to the actual flow (see Henney et al. 2005) and is a common assumption made in connecting the H$^+$ region to PDR (Carral et al. 1994; Petuchowski & Bennett 1993). The start of the PDR is defined to be the position where $n(H^+)/n(H_{tot})$ falls below 1%. Given the initial conditions in the H$^+$ region and an equation of state, the resulting emission-line spectrum is then computed self-consistently.

# 4 Results

## 4.1 Percentage of [C II] Emission from the H$^+$ Region

The calculation results for all $U$, $n_{H^+}$, $T_*$, and stellar continuum are shown in Figures 1-5. These figures show the percentage of [C II] emission that comes from the H$^+$ region, $[C\ II]^{\%}_{H^+} = 100 \times \frac{[C\ II]_{H^+}}{[C\ II]_{H^+} + [C\ II]_{PDR}}$. In general, the lower the density in the H$^+$ region, the higher $[C\ II]^{\%}_{H^+}$. Figures 1-5 show that the H$^+$ region must be included when trying to calculate the [C II] emission. For most of parameter space considered by our calculations, $[C\ II]^{\%}_{H^+}$ exceeds 10% and can be as high as 50% for low-density H$^+$ regions.

Although $n_{H^+}$ is the primary factor in determining $[C\ II]^{\%}_{H^+}$, the stellar continuum also plays a role. For $T_*$=30,000 K (Figure 1), few photons capable of ionizing C$^+$ ($h\nu > 24.38$ eV) are produced. This leaves C$^+$ as the dominant stage of ionization throughout most of the H$^+$ region. Therefore, for low $T_*$ and a constant $n_{H^+}$, $[C\ II]^{\%}_{H^+}$ increases with increasing $U$. However, what constitutes "low $T_*$" depends on the choice of stellar continuum. The Kurucz stellar continuum produces fewer ionizing photons than either a WMBasic or a blackbody continuum. The Kurucz continuum, therefore, requires higher $T_*$ in order to produce a significant $C^{2+}$ zone in the $H^+$ region. This is seen in Figure 2, where the Kurucz continuum shows little difference between Figures 1 and 2, while the WMBasic and blackbody continuum begins to form $C^{2+}$ in the $H^+$ region, leading to less $[C\ II]^{\%}_{H^+}$ for the same set of $U$ and $n_{H^+}$. In general, WMBasic and blackbody continua produce very similar results.

For high values of $T_*$, all three stellar continua produce significant $C^{2+}$ in the $H^+$ region. The primary factor determining $[C\ II]^{\%}_{H^+}$ then becomes (besides $n_{H^+}$) the size of the $H^+$ region relative to the size of the $C^+$ region in the PDR. The larger the size of the $H^+$ region is relative to the [C II] emitting region in the PDR, the larger $[C\ II]^{\%}_{H^+}$ becomes. The relative size of each region depends on the hydrogen ionizing flux and $G_0$. The diagonal contours in Figures 4 and 5 closely follow the flux of photons (since, for a given $T_*$, $U = \dfrac{\phi}{n_e c} \propto \dfrac{G_0}{n}$ -- see Figure 27 of Abel et al. 2005). For low $G_0/n$, the $C^+ \rightarrow C^0 \rightarrow CO$ transition occurs at a smaller $A_V$, leading to less $C^+$ column density and lower gas temperature in the PDR (Kaufman et al. 1999). Both of these suppress the [C II] emission from the PDR and increase the relative importance of the $H^+$ region.

The calculations shown here are only for a single metallicity, which can change the predicted [C II] emission from $H^+$ regions. Kaufman, Hollenbach, & Wolfire (2006, in preparation) note that emission line intensities from metals will, to first order, linearly increase with metallicity.

## 4.2 Correlation between [C II] and [N II] 205 μm Intensity

Figure 6 shows the correlation of [C II] emission, both from the $H^+$ region and $H^+$ + PDR, with [N II] 205 μm emission for all calculations. Our calculations show a strong correlation between [N II] and [C II] from the $H^+$ region. Physically, this correlation is due to two reasons. One is [N II] and [C II] have similar critical densities (80 and 50 cm$^{-3}$; respectively, Osterbrock & Ferland 2005). This means that both lines will become collisionally de-excited at about the same $H^+$ density. The other reason is, in $H^+$ regions, there is significant overlap between the $C^+$ and $N^+$ zones (Figure 7).

Our calculations reproduce the observed correlations found by Petuchowski & Bennett (1993). This is somewhat surprising, considering their observations of the WIM, and our calculations are for $H^+$ regions adjacent to PDRs. It seems, therefore, that Figure 6 is robust. The solid line indicates the power-law fit to our calculations. This fit is given by the equation:

$$Log[I^{CII}_{H^+}] = 0.937 Log[I^{NII}_{H^+}] + 0.689 \quad (\text{erg cm}^{-2}\ \text{s}^{-1}) \tag{1}$$

where $I^{CII}_{H^+}$ and $I^{NII}_{H^+}$ are the intensities of [C II] and [N II] from the H$^+$ region. Equation 1 agrees well with the equation derived by Heiles (1994) from analytical considerations and the Petuchowski & Bennett (1993) WIM data (equation 12a and the dotted line in Figure 6).

The effects of metallicity on the [C II]-[N II] correlation was also investigated. The same input parameters were used as before, only this time with the metallicity (and grain abundance) reduced by 1 dex. Additionally, since Figures 1-5 show little variation between $[C\ II]^{\%}_{H^+}$ predictions from WMBasic or a blackbody continuum, the low metallicity calculations shown here use a blackbody spectrum.

The results of varying metallicity are shown in Figure 6. Reducing the metallicity reduces the C$^+$ abundance in the H$^+$ region, thereby reducing the emission. In the PDR, the effects of metallicity are not as significant. Even though the C$^+$ abundance is reduced by 1 dex, the C$^+$ column density increases, since dust is less effective in attenuating the UV radiation field. This leads to less [C II] emission from the H$^+$ region. The best-fit to the Z=0.1×Z$_{orion}$ calculations is therefore lower than for Z=Z$_{orion}$. The best-fit is given by:

$$Log[I^{CII}_{H^+}] = 0.850 Log[I^{NII}_{H^+}] + 0.130 \quad (\text{erg cm}^{-2}\ \text{s}^{-1}) \qquad (2)$$

It is worth noting that equation 1 depends on the abundance of C$^+$ and N$^+$ in the H$^+$ region. The calculations presented here consider all ionization stages of C and N and therefore takes into account the abundance of all other ionization states. Equation 1 does assume an overall C/N abundance ratio (~4.3). To take into account C/N abundance variations, one should scale equation 1 and 2 under the assumption that $I^{CII}_{H^+}$ and $I^{NII}_{H^+}$ scale linearly with abundance.

## 4.3 Application to Observation

Figures 1-5 and equation 1 have important practical application to observation. Infrared mission such as SOFIA and Herschel will commonly observe both emission lines in a single pointing. Equations 1 & 2 therefore provides a simple way to subtract out $I^{CII}_{H^+}$ before applying standard PDR computational techniques. Additionally, SOFIA and Herschel will have sufficient spectral resolution that it should be able to resolve the broad, H$^+$ region emission from cooler PDR emission.

If other spectral diagnostics are observed that derive $n_e$, $U$, and $T_*$, then Figures 1-5 offer a way to determine the equation of state, the relationship between density, temperature, and pressure in a star-forming region. By deriving $n_e$, $U$, and $T_*$, an estimate of $I^{CII}_{H^+}$ can be made if the metallicity and C/N abundance ratio is known. Observations with SOFIA or Herschel will tell us $[C\ II]_{H^+}^{\%}$. The question is what equation of state can reproduce the observed $[C\ II]_{H^+}^{\%}$. If the assumption of constant pressure is correct, then Figures 1-5 should reproduce the observations. If not, then the equation of state is influenced by other dynamical processes such as shocks, advection, or magnetic fields. SOFIA or Herschel observations of a large sample of $H^+$ regions with adjacent PDRs, combined with theoretical calculations, provides one way to probe the equation of state.

# 5 Conclusions

1. This paper has calculated the C II emission from $H^+$ regions over a wide range of $T_*$, $U$, $n_e$, and stellar continuum shapes under the assumption of constant pressure. For most conditions, at least 10% and sometimes up to 50%-60% of the total [C II] emission can come from ionized gas.

2. The primary factors in determining the [C II] contribution to $H^+$ region are the electron density, the size of the $H^+$ region relative to the size of the $C^+$ region in the PDR, and the stellar continuum. The electron density controls the collisionally excitation of 158 μm line, the relative size of each region determine over what physical extent [C II] emission can occur, and the stellar continuum controls transition from $C^{2+}$ to $C^+$ and therefore the physical extent of the [C II] emitting region in the $H^+$ region.

3. The calculations presented here reproduce the observed correlation between [N II] and [C II], and an equation relating [N II] emission with [C II] was derived. This equation deviates only slightly from the equation given in Heiles (1994), and is therefore applicable to the WIM as well as typical $H^+$ regions adjacent to PDRs.

4. Future observations, combined with the [N II] → [C II] correlation and theoretical calculations, can determine the equation of state linking the $H^+$ region to the PDR. This has important consequences to the study of dynamical processes and the role of magnetic fields in star-forming regions.

Acknowledgements: I would like to acknowledge the anonymous referee for useful comments on earlier drafts of this work. I would also like to acknowledge the support of the Center for Computational Sciences at the University of Kentucky, along with Michael Kaufman and Gary Ferland for comments regarding this manuscript.Acknowledgements: I would like to acknowledge the anonymous referee for useful comments on earlier drafts of this work. I would also like to acknowledge the support of the Center for Computational Sciences at the University of Kentucky, along with Michael Kaufman and Gary Ferland for comments regarding this manuscript.

# 7 Figures

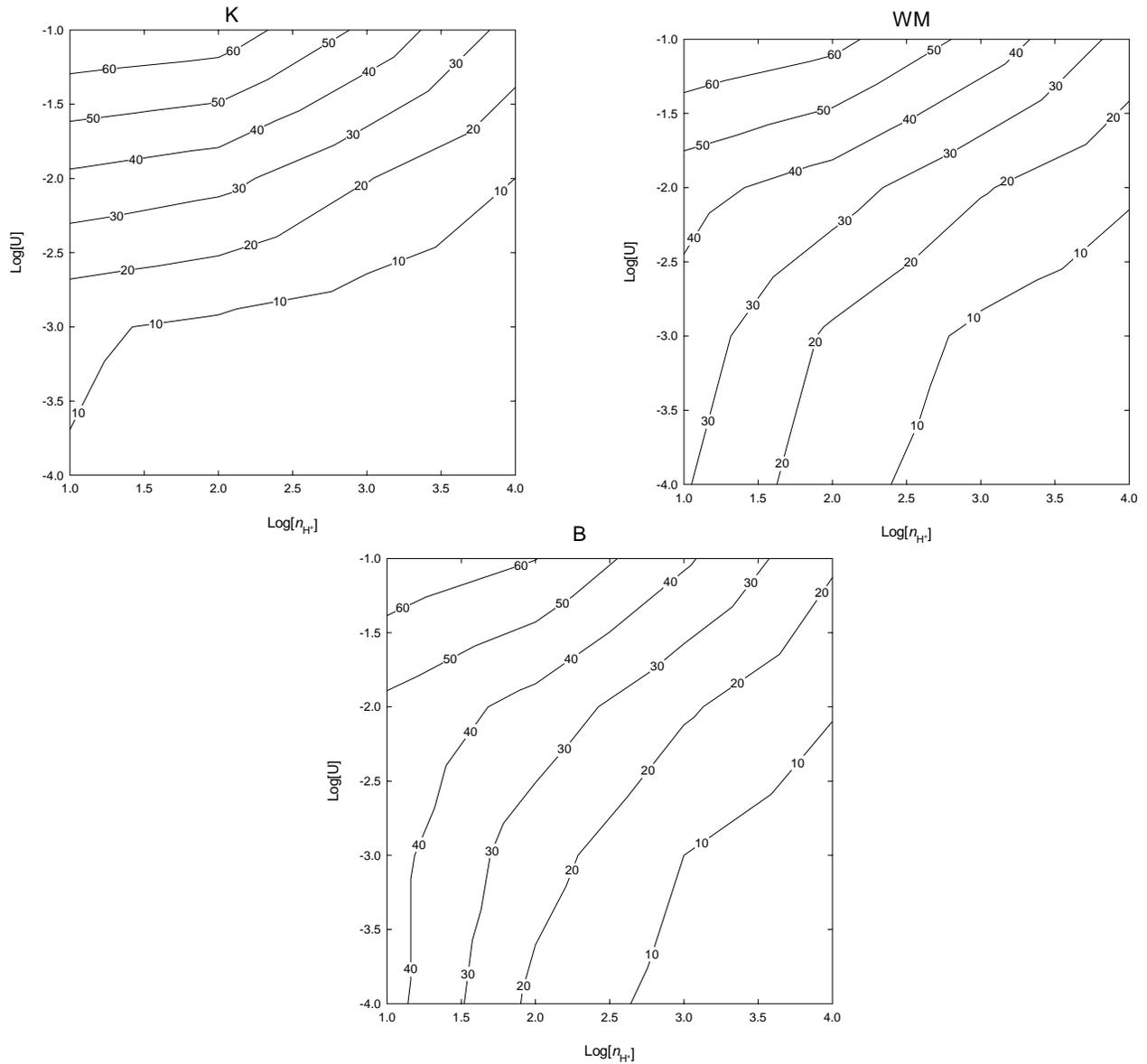

Figure 1 $[C\ II]^{\%}_{H^+}$ for 30,000 K for the Kurucz (K), WMBasic (WM) and blackbody (B) stellar continua. When $T_*$ is low, as is the case here, $C^+$ is the dominant ionization stage in the $H^+$ region as well as the PDR. This allows for substantial [C II] emission in the $H^+$ region.

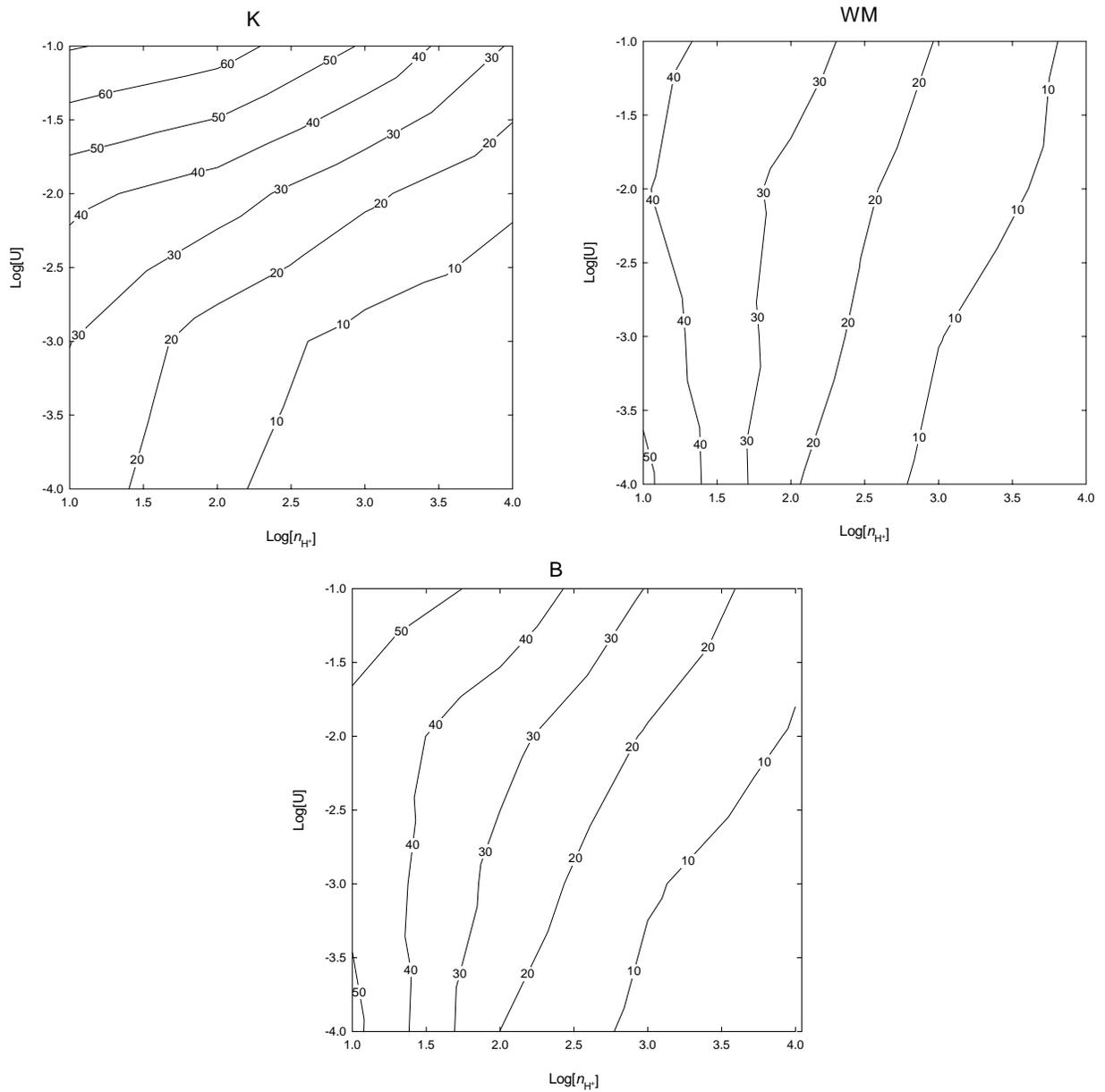

Figure 2 $[C\ II]^{\%}_{H^+}$ from $H^+$ for 35,000 K for the Kurucz (K), WMBasic (WM) and blackbody (B) stellar continua. For the Kurucz stellar atmosphere, $T_*$ is still too low to produce large amounts of $C^{2+}$, leading to $C^+$ being the dominant ionization state in the $H^+$ region. The WMBasic and Blackbody continuum produces more energetic photons than Kurucz, increasing the $C^{2+}$ abundance. For these two continua, the contours are more vertical, indicating the dependence of [C II] emission on the electron density in $H^+$ regions.

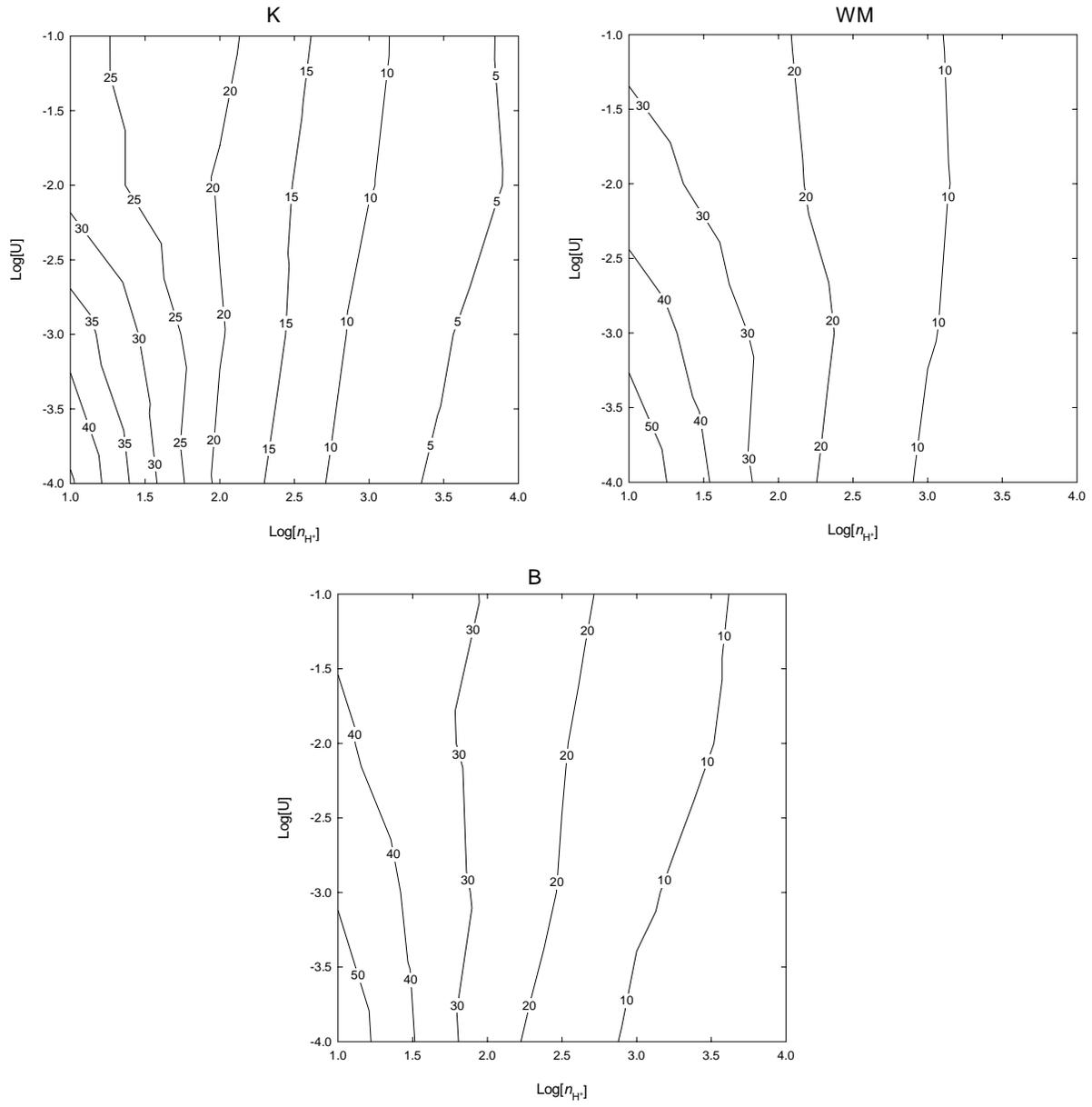

Figure 3 $[C\ II]^{\%}_{H^+}$ from H+ for 40,000 K for the Kurucz (K), WMBasic (WM) and blackbody (B) stellar continua. The contours are approximately vertical, indicating the dependence of [C II] emission on the electron density in H+ regions.

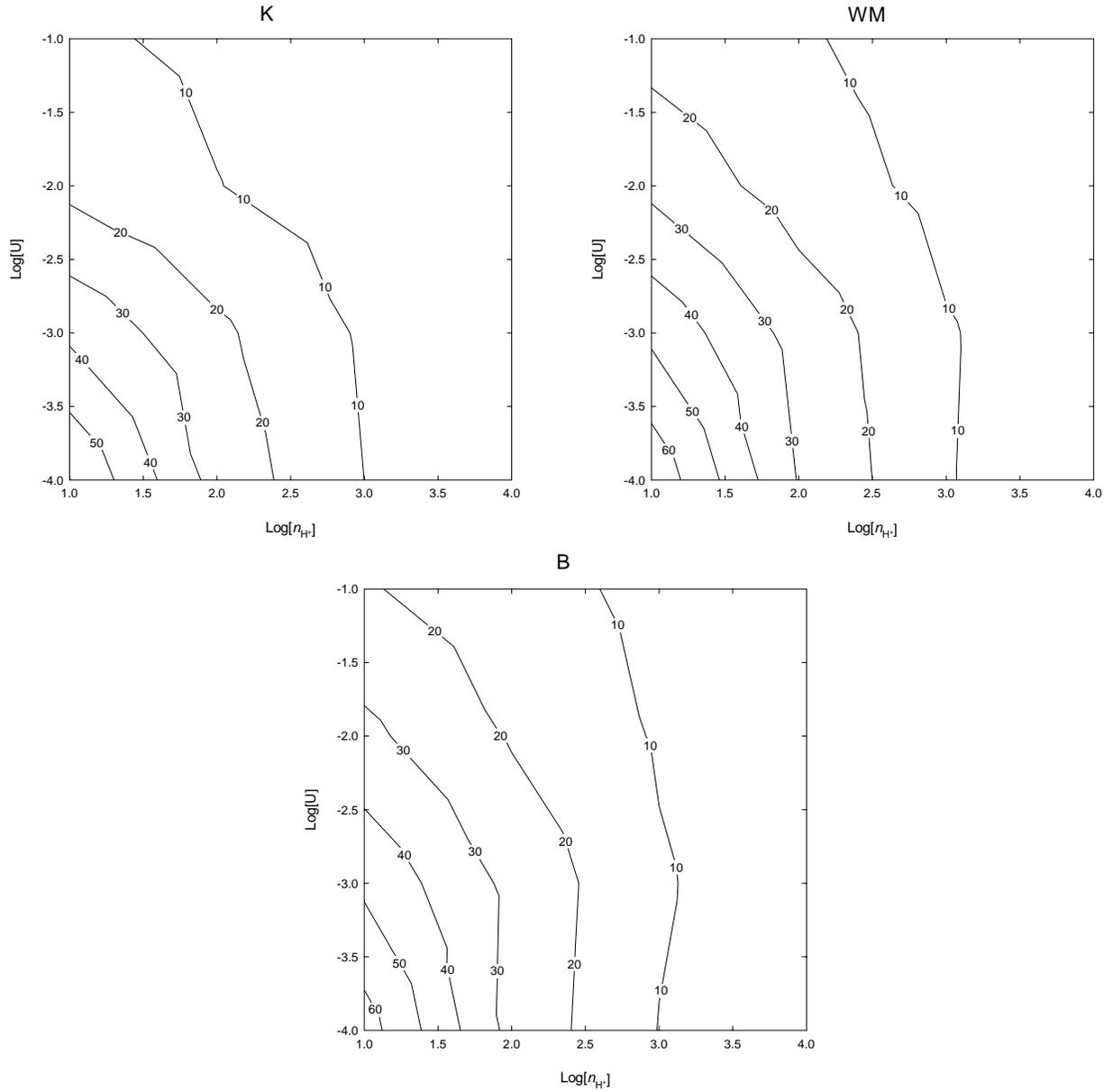

Figure 4 $[C\ II]^{\%}_{H^+}$ from H$^+$ for 45,000 K for the Kurucz (K), WMBasic (WM) and blackbody (B) stellar continua. For low-density H$^+$ regions, more [C II] emission is produced from the H$^+$ region. Additionally, there is a dependence on the relative size of the H$^+$ region and C$^+$ region in the PDR. For low $U$, low $n_{H^+}$ regions, G$_0$ is ~1-10 (Abel et al. 2005). This leads to a smaller C$^+$ region in the PDR, and hence less [C II] emission from the PDR.

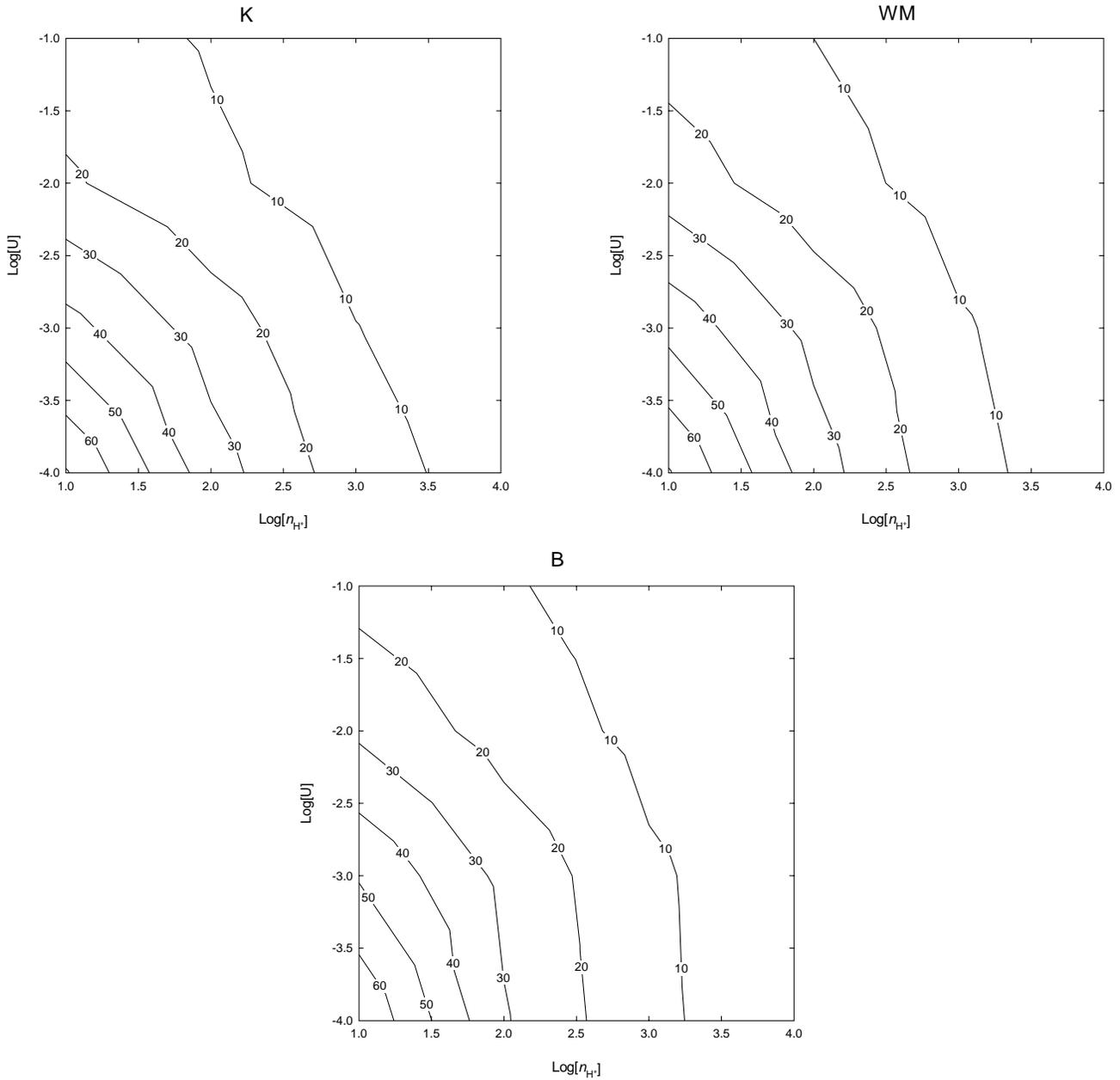

Figure 5 $[C\ II]^{\%}_{H^+}$ from H$^+$ for 50,000 K for the Kurucz (K), WMBasic (WM) and blackbody (B) stellar continua. For low-density H$^+$ regions, more [C II] emission is produced from the H$^+$ region. Additionally, there is a dependence on the relative size of the H$^+$ region and C$^+$ region in the PDR. For low $U$, low $n_{H^+}$ regions, G$_0$ is ~1-10 (Abel et al. 2005). This leads to a smaller C$^+$ region and hence less [C II] emission from the PDR.

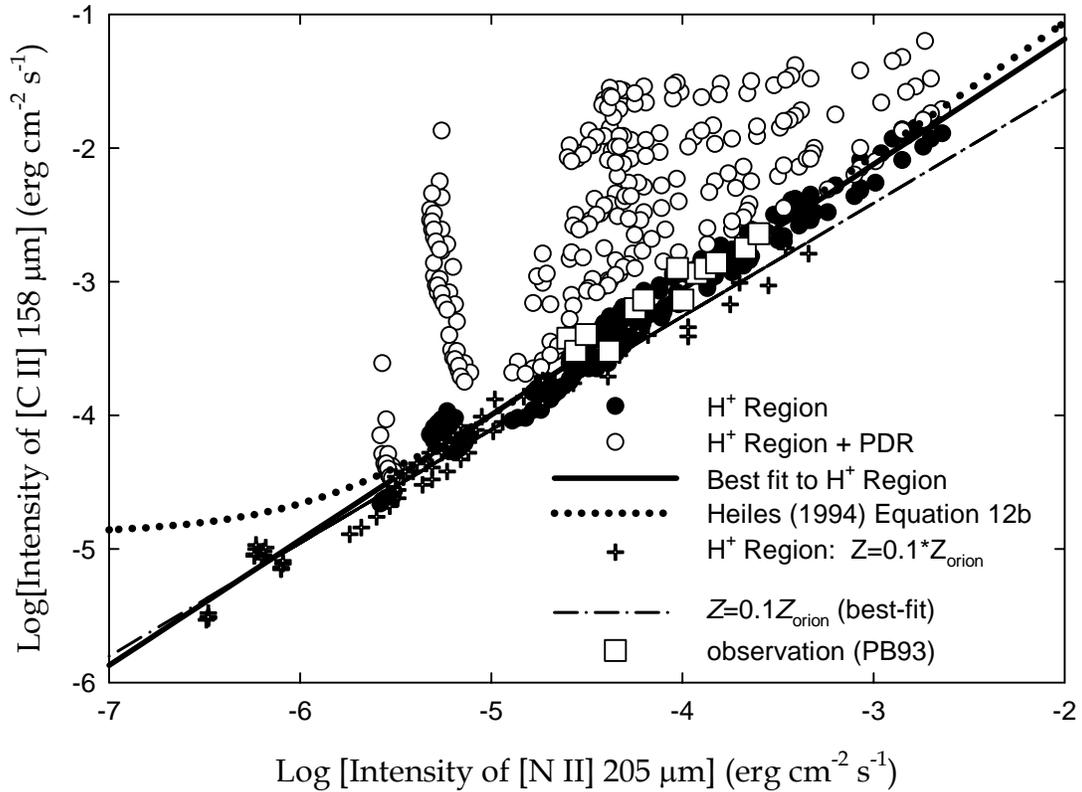

Figure 6 The correlation between [N II] and [C II] emission. The filled circles represent the contribution of [C II] from the H+ region, while the open circles represent the total [C II] emission (H+ + PDR). The open squares represent the COBE observations taken from Petuchowski & Bennett (1993—referred to in the figure as PB93). The solid line represents the best-fit equation to our data, with the equation of the line given in equation 1. The dotted line is the fit from Heiles (1994). One can easily see the strong correlation between [N II] and [C II] emission from the H+ region, with equation 1 having a correlation coefficient $R^2$ of 0.97. Also shown is the effect of reducing the metallicity to $Z=0.1Z_{orion}$. Lowering the metallicity lowers $[C\ II]^{\%}_{H^+}$, since H+ region emission line intensities scale with metallicity. The [C II] emission from PDRs, however, depends on the total column density of $C^+$, which is less sensitive to changes in $Z$ (Kaufman et al. 1999; Kaufman, Hollenbach, & Wolfire 2006).

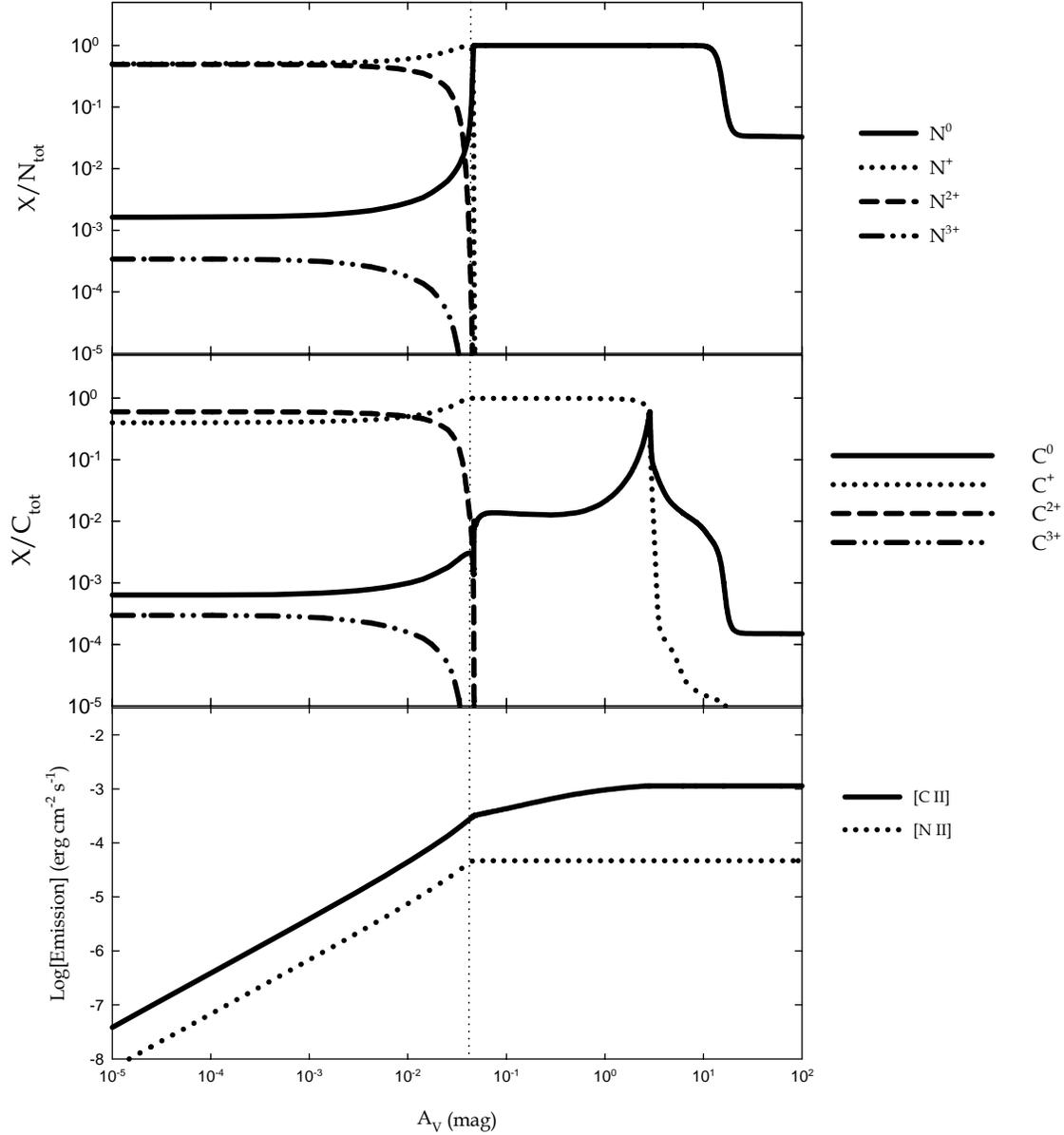

Figure 7 Carbon and nitrogen ionization structure vs. $A_V$ for a 40,000 K Blackbody with $U = 10^{-3}$ and $n_{H^+} = 10^2$ cm$^{-3}$. Also shown is the [C II] and [N II] emission vs. $A_V$. The vertical dotted line represents the hydrogen ionization front. The abundance of C$^+$ in the H$^+$ region is significant (~40%). Close to the I-front, C$^+$ is the dominant ionization stage, with N$^+$ following the same structure. The resulting correlation between [C II] and [N II] emission in the H$^+$ region is clearly seen. In the PDR, N → N$^0$, while C$^+$ remains the dominant stage of ionization, which allows for further [C II] emission. Overall, ~25% of the [C II] emission arises in the H$^+$ region. The drop in N$^0$, C$^+$, and C$^0$ deep in the calculation is due to the formation of N$_2$ and CO (not shown here).